\documentclass[runningheads,a4paper]{llncs}

\usepackage[T1]{fontenc}

\usepackage{amssymb,amsmath,bbm}
\usepackage{graphicx}

\usepackage{url}
\urldef{\mailjr}\path|juste.raimbault@polytechnique.edu|

\newcommand{\keywords}[1]{\par\addvspace\baselineskip
\noindent\keywordname\enspace\ignorespaces#1}

\newcommand{\noun}[1]{\textsc{#1}}

\begin{document}

\mainmatter  

\title{An Applied Knowledge Framework to Study Complex Systems}

\titlerunning{An Applied Knowledge Framework}

%
%
\author{\noun{Juste Raimbault}$^{1,2}$}
\authorrunning{An Applied Knowledge Framework}

\institute{$^{1}$ UMR CNRS 8504 G{\'e}ographie-Cit{\'e}s, Paris, France\\
$^{2}$ UMR-T IFSTTAR 9403 LVMT, Champs-sur-Marne, France\\
\mailjr
}

\toctitle{Lecture Notes in Computer Science}
\tocauthor{Authors' Instructions}
\maketitle

\begin{abstract}
The complexity of knowledge production on complex systems is well-known, but there still lacks knowledge framework that would both account for a certain structure of knowledge production at an epistemological level and be directly applicable to the study and management of complex systems. We set a basis for such a framework, by first analyzing in detail a case study of the construction of a geographical theory of complex territorial systems, through mixed methods, namely qualitative interview analysis and quantitative citation network analysis. We can therethrough inductively build a framework that considers knowledge entreprises as perspectives, with co-evolving components within complementary knowledge domains. We finally discuss potential applications and developments.

\end{abstract}

\section{Introduction}

The understanding of processes and conditions of scientific knowledge production are still mainly open questions, to which monuments of epistemology such as Kant's Critique of Pure Reason, and more recently Kuhn's study of ``the structure of scientific revolutions''~\cite{kuhn1970structure} or Feyerabend's advocacy for a diversity of viewpoints \cite{feyerabend1993against}, have brought elements of answer from a philosophical approach. A more empirical point of view was brought also recently with quantitative studies of science, in a way a \emph{quantitative epistemology} that goes far beyond rough bibliometric indicators~\cite{cronin2014beyond}. Contributions harnessing complexity, i.e. studying complex systems in a very broad sense, can be shown to have produce very diverse frameworks that can be counted as building bricks contributing to answers to the above high-level question. We will in the following use the term Knowledge Framework, for any such framework having an epistemological component tackling the question of nature of knowledge or knowledge production. To illustrate this, we can mention such frameworks in different domains, at different levels and with different purposes. For example, \cite{durantin2017disruptive} explores the potentialities of coupling engineering with design paradigms to enhance disruptive innovation. Also in Knowledge Management, using the constraint of innovation as an advantage to understand to complex nature of knowledge, \cite{carlile2004transferring} introduces knowledge domains boundaries and production processes. Also introducing a meta-framework, but in the field of system engineering, \cite{gemino2004framework} recommends to use grammars to compare Conceptual Modeling Techniques. Meta-modeling frameworks can also be understood as Knowledge Frameworks. \cite{cottineau2015modular} describes a multi-modeling framework to test hypotheses in simulation of socio-technical complex systems. \cite{golden2012modeling} postulates a unified formulation of systems, including necessarily different types of knowledge on a system on its different description components.
 
A possible explanation for this richness is the fundamental reflexive nature of the study of Complex Systems: because of the higher choice in methodology and what aspects of the system to put emphasis on, a significant part of a modeling or design entreprise is an investigation at a meta-level.  Furthermore, studies of knowledge production are mainly rooted in complexity, implying a reflexive nature of theories accounting knowledge on complexity, as Hofstadter had well highlighted in \cite{hofstadter1980godel} by noticing the importance of ``strange loops'', i.e. feedback loops allowing reflexivity such as a theory applying to itself, in what constitutes intelligence and the mind. Artificial intelligence is indeed a crucial field regarding our issues, as its progresses imply a finer understanding of the nature of knowledge. \cite{2017arXiv170401407M} introduces a meta-framework for a general typology of approaches in Artificial Intelligence, what is a Knowledge Framework not in the proper sense but in a specific applied case.

The level of frameworks described above may be very general but is conditioned to a certain field or discipline, and to a certain approach or methodology. There exists to our knowledge no framework realizing a difficult exercise, that is to capture a certain structure of knowledge production at an epistemological knowledge, but conjointly is thought in a very applied perspective, with direct consequence in the design and management of complex systems. The contribution of this paper attempts to set a basis for a Knowledge Framework realizing this in the case of Complex Systems. To perform that, we postulate that the tension between these two contradictory objective is an asset to avoid on one side an impossible overarching generality and on the other side a too restraining domain-specific specificity. Based on the idea of complementary Knowledge Domains introduced by~\cite{livet2010}, its central aspect is a cognitive approach to science inducing co-evolutive processes of knowledge domains and their carriers. A first sketch of this framework was presented by~\cite{raimbault:halshs-01505084}, in the specific case of complex territorial systems as studied by theoretical and quantitative geography. We choose to introduce it here with an inductive approach, i.e. starting from a concrete case study that has mainly inspired the construction of the framework to end with its generic description.

The rest of the paper is organized as follows : the next section develops case studies, more precisely a detailed study of a geographical theory of complex urban systems, and a short example from engineering to illustrate the transferability of concepts. The third section specifies the definitions and formulates the epistemological framework. We finally discuss issues on applicability, and potential developments such as a mathematical version of the framework.

\section{Case Studies}

\subsection{Genesis of the Evolutive Urban Theory}

The first case study relates the construction of the \emph{Evolutive Urban Theory}, a geographical theory considering territorial systems from a complexity perspective, that have been developed for around 20 years. We analyse its genesis using mixed methods, namely semi-directed interviews with main contributors, and quantitative bibliometric analysis of main publications. Interviews were done following methodological standards \cite{legavre1996neutralite} to ensure a limited interference of the interviewer's experiences but not make it fully disappear to ensure a precise context enhancing the fluency of the interviewed. We use here interviews\footnote{Both have a length of around 1h. Sound and transcript text are available under a CC Licence at \texttt{https://github.com/JusteRaimbault/Interviews}~\cite{raimbault2017entretiens}. Interviews are in French and translations here are done by the author.} with Pr. D. Pumain who introduced and developed mainly the theory, and Dr. R. Reuillon, whose research on intensive and distributed computation and model exploration has been a cornerstone of latest developments.

Let first give an overview of its content. This theory was first introduced in~\cite{pumain1997pour} which argues for a dynamical vision of city systems, in which self-organization is key. Cities are interdependent evolutive spatial entities whose interrelations produces the macroscopic behavior at the scale of city system. The city system is also described as a network of city what emphasizes its view as a complex system. Each city is itself a complex system in the spirit of~\cite{berry1964cities}, the multi-scale aspect being essential in this theory, since microscopic agents convey system evolution processus through complex feedbacks between scales. The positioning within Complex System Sciences was later confirmed~\cite{pumain2003approche}. It was shown that this theory provide an interpretation for the origin of pervasive scaling laws, resulting from the diffusion of innovation cycles between cities~\cite{pumain2006evolutionary}. The aspect of resilience of system of cities, induced by the adaptive character of these complex systems, implies that cities are drivers and adapters of social change~\cite{pumain2010theorie}. Finally, path dependance yield non-ergodicity within these systems, making ``universal'' interpretations of scaling laws difficultly compatible with evolutive urban theory~\cite{pumain2012urban}. The construction of models of urban systems has been a key component for the theory, starting with the first Simpop model~\cite{sanders1997simpop}. Later example include for example the Simpop2 model, an agent-based model taking into account economic processes, that simulates growth patterns on long time scales for Europe and the United States~\cite{doi:10.1177/0042098010377366}. The latest accomplishment of the evolutive theory relies in the output of the ERC project GeoDivercity, presented in~\cite{pumain2017urban}, that include both advanced technical (software OpenMole\footnote{http://openmole.org/}~\cite{reuillon2013openmole}), thematic (knowledge from SimpopLocal~\cite{schmitt2014modelisation} and Marius models~\cite{cottineau2014evolution}) and methodological (incremental modeling~\cite{cottineau2015incremental}) progresses.

The striking feature in the construction of all this is the balance between the different \emph{types} of knowledge, of which a typology will be the starting point of our construction. The relation between theoretical considerations and empirical cases studies is fundamental. Indeed, the seminal article \cite{pumain1997pour} is already positioned as an ``advocacy for a less ambitious theory, but that does not neglects the back-and-forth with observation''\footnote{page 2, trad. author}. We shall now turn to interviews to better understand the implications of the intrication of types of knowledge. D. Pumain traces back germinal ideas back to her graduate student work in 1968, when ``everything started with a question of data''. The interest for cities, and \emph{change in cities}, was driven by the availability of a refined migration flow dataset at different dates. Also rapidly, ``[they] were frustrated that methods were missing'', but the access to the computation center (\emph{technical tool}) allowed the test of newly introduced methods and models, linked to the Prigogine approach to complexity. Methods were however still limited to grasp the heterogeneity of spatial interactions. A progressively specified need and a chance encounter, with ``a lady working on neural networks and agent-based modeling at the Sorbonne'', led to a bifurcation and a new level of interaction between modeling, theory and empirical knowledge: in 1997, two seminal articles, one stating the theoretical basis and the other introducing the first Simpop model, were published. From this point, it was clear that all modeling entreprise was conditioned to empirical knowledge of geographical case studies and theoretical assumptions to test. Methods and technical tools took also a necessary role, when specific model exploration methods were developed together with the Software OpenMole. R. Reuillon relates that a qualitative shift of knowledge was rapidly made possible when systematic model exploration methods were introduced to understand the behavior of the SimpopLocal model. Initially, geographers were not sure if the model worked at all, in the sense that it produced expected stylized facts such as the emergence of hierarchy in a system of cities. Satisfying trajectories were found for some parameter values through genetic algorithm calibration, with distributed computation on grid~\cite{schmitt2014half}. The existence of multiple candidate solutions for parameter values is a barrier for concrete questions of necessity or sufficiency of a given mechanism of the agent-based model. This need, coming from the domain of empirical and theoretical geographical knowledge, led to the design of a specific algorithm the calibration profile, which a methodological advance in model exploration~\cite{reuillon2015}. This virtuous circle was continued with the Marius model family~\cite{cottineau2014evolution} and the Parameter Space Exploration algorithm~\cite{10.1371/journal.pone.0138212}. R. Reuillon evaluates its impact from a Computer Scientist point of view: ``I'm not sure if [geographers] were immediately conscious of the amplitude of the result, that was really heavy, people working with us directly saw it.'' This positive vision is confirmed by D. Pumain, who highlights the benefits of these new methods for geographical knowledge, and that it was the first time that research led to publications at the edge of knowledge both in geography and computer science.

Taking a step back, emerges a typology of domains in which knowledge was created but also necessary for the other domains in the genesis of the Evolutive Urban Theory. The collection of data and construction of datasets is a first requirement for any further knowledge. From data are extracted empirical stylized facts, from which are induced theoretical hypotheses. Theory can then be tested for falsification, in the empirical domain but also through models, for example by doing targeted experiments in models of simulation. New methods are developed to better explore them. Tools are crucial at each step, to implement model, do data mining for example or collect and format data for example. The previous analysis reveals how these domains are interdependent, are in a sense \emph{co-evolutive}.

\begin{figure}[h!]
\hspace{-4cm}
\includegraphics[width=1.5\textwidth]{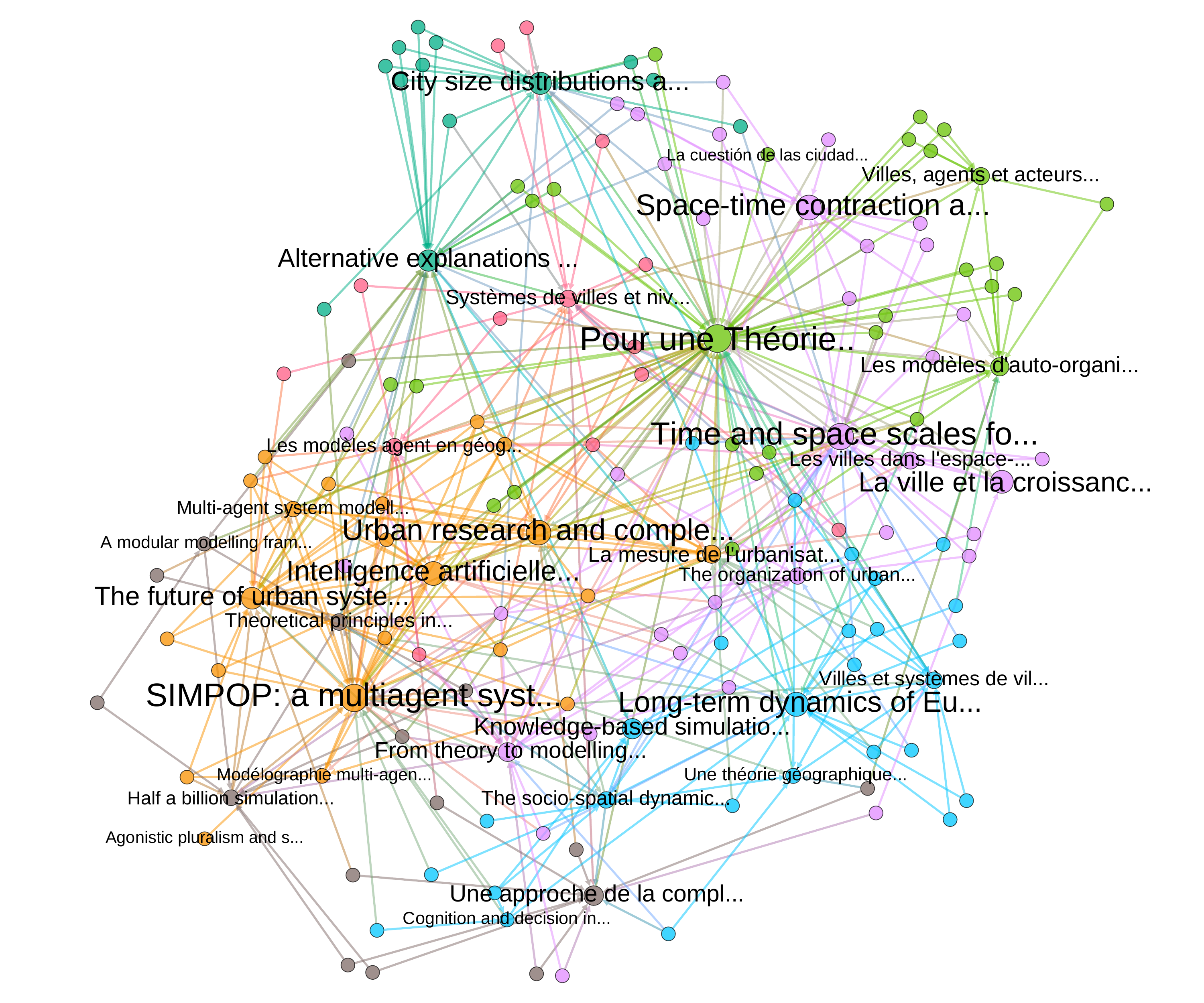}
\caption{\textbf{Citation Network of main publications of Evolutive Urban Theory.} The network is constructed the following way: starting from the two seminal publications~\cite{pumain1997pour} and \cite{sanders1997simpop}, we get citing publications, filter conditionally to one of the main contributors, get again citing publications and filter. Nodes are publications ($\left|V\right|=155$), the size corresponding to eigenvector centrality, and edges are directed citation links ($\left|E\right|=449$). Colors are communities obtained with Louvain clustering algorithm (7 communities, modularity 0.39).}
\label{fig:citnw}
\end{figure}

We back up now this qualitative analysis with a modest quantitative bibliometric analysis. The idea is to investigate the structure of the core citation network of main publications constructing the Evolutive Urban Theory. We construct the citation network as described in Fig.~\ref{fig:citnw}, by using the data collection tool provided by~\cite{raimbault2016indirect}\footnote{all code and data are available at \\\texttt{https://github.com/JusteRaimbault/CityNetwork/tree/master/Models/QuantEpistemo}}. Starting from the two seminal publications \cite{pumain1997pour} and \cite{sanders1997simpop}, the backward citation network is obtained at depth 2 (references citing these initial references, and the ones citing the citing), with filtering for the first step on authors to have at least one main contributor of the Theory (that we take as \emph{Pumain}, \emph{Sanders} and \emph{Bretagnolle}, according to the full Pumain's interview). We remove nodes of degree 1, to have the core structure only of the ego network. Note that we do not have missing links between nodes at the first level, because all citing links were retrieved. Network has a density of 0.019, what is rather high for a citation network, and the signature of a high level of dependency between publications. Starting from two separate nodes, we could have in theory distinct connected components, but as expected the network has only one because both aspects are strongly interconnected. To analyse the structure in a finer way, we detect communities using Louvain clustering algorithm, and evaluate the directed modularity of the partition as described by \cite{nicosia2009extending}. We show in Fig.~\ref{fig:citnw} a visualization of the network. We obtain 7 communities with a modularity value of 0.39. To ensure the significance of modularity, we proceed to Monte Carlo simulations and randomize citation links 100 times, computing each time the modularity of communities within the randomized network. We obtain an average directed modularity of $\bar{m} = 0.002 \pm 0.015$, making the modularity of the real network highly significant (more than 200 standard deviations). We analyse the content of communities by looking at publications of the first level. We find that communities are roughly consistent with the typology of domains: one on methods, three on spatio-temporal modeling of urban systems that mixes empirical and modeling, one conceptual, one on Simpop models, and a last on scaling laws that is fully empirical. Data papers are not yet current practice in geography and specific papers tackling the Data domain can be found in the network. An increased citation rate between papers of the same domain could be expected because of the scientific standard to always situate a contribution regarding similar works. The significant value of modularity confirms that domains are consistent regarding an certain endogenous structure of knowledge production.

\subsection{Engineering the Metropolitan}

After the glance on domains of knowledge extracted in the previous case study, we propose to take the corresponding point of view on a rather different example more related to technology and engineering. We interpret thus issues of engineering related to Parisian metropolitan system through this prism of Knowledge Domains. Taking the example of the progressive automatization of line 1, considered widely as a technical achievement, several integrated empirical and modeling studies were preliminary conducted~\cite{belmonte2008automatisation}. The use and adaptation of particular methods such as agent-based modeling is crucial for the development of innovative autonomous transportation~\cite{balbo2016positionnement}. In this engineering problem, some technical solutions such as platform doors may be seen as tools that also evolve, and are necessary for a new conceptual approach (\emph{automatic transportation}) to be implemented~\cite{foot2005faut}. But they may also have interactions with other aspects of conceptual knowledge, such as management and organisation within the operator~\cite{foot1994ratp}. The complex multi-dimensional aspect of innovation for such systems was already highlighted for a while as~\cite{hatchuel1988stations} shows. Other technical aspects, such as civil engineering issues~\cite{moreno2016etude}, are also put in line when developing such a new approach, and they necessitate at least empirical and modeling, if not more, Knowledge Domains. This rather short example is an illustration of how the interpretation of knowledge domains can be applied to the engineering and management of a complex industrial systems. Specific details would be needed for a more in-depth application, but we claim to have a proof-of-concept here.

\section{Knowledge Framework}

We can formulate now inductively the knowledge framework. As mentioned, it takes the idea of interacting domains of knowledge from the framework introduced by~\cite{livet2010}, but extends these domains and takes a novel epistemological position, focusing on co-evolutive dynamics of agents and knowledge.

\paragraph{Constraints}

To be particularly fitted for the study and management of complexity, we postulate that the framework must meet certain requirements, especially to take into account and even favor the \emph{integrative nature of knowledge}, as illustrated by the importance of interdisciplinarity and diversity in the case studies. The framework must thus be favorable to the following:
\begin{itemize}
\item Integration of disciplines, as Complex Systems are by essence at the crossing of multiple fields
\item Integration of knowledge domains, i.e. that no particular type of knowledge must be privileged in the production process\footnote{this is not incompatible with very strict system specifications, as multiple paths are possible to obtain the same fixed final state}
\item Integration of methodology types, in particular breaking the artificial boundaries between ``quantitative'' and ''qualitative'' methods, which are particularly strong in classical social sciences and humanities.
\end{itemize}

\paragraph{Epistemological Fundations}


Our epistemological positioning relies on a cognitive approach to science, given by Giere in~\cite{giere2010explaining}. The approach focuses on the role of cognitive agents as carriers and producers of knowledge. It has been shown to be operational by \cite{giere2010agent} that studies an agent-based model of science. These ideas converge with Chavalarias' Nobel Game~\cite{chavalarias2016s} that tests empirically the balance between exploration and falsification in the collective scientific enterprise. This epistemological positioning has been presented by Giere as \emph{scientific perspectivism}~\cite{giere2010scientific}, which main feature is to consider any scientific entreprise as a \emph{perspective} in which \emph{agents} use \emph{media} (models) to represent something with a certain purpose. To make it more concrete, we can position it within Hacking's ``check-list'' of constructivism~\cite{hacking1999social}, a practical tool to position an epistemological position within a simplified three dimensional space which dimensions are different aspects on which realist approaches and constructivist approach generally diverge: first the contingency (path-dependency of the knowledge construction process) is necessary in the pluralist perspectivist approach, secondly the ``degree of constructivism'' is quite high because agents produce knowledge, and finally the stability of theories depends on the complex interaction between the agents and their perspectives. It was presented for these reasons as an intermediate and alternative way between absolute realism and skeptical constructivism~\cite{brown2009models}. The \emph{perspective} plays a central role in the framework.

\paragraph{Knowledge Domains}

We postulate the following knowledge domains, with their definitions:
\begin{itemize}
\item \textbf{Empirical.} Empirical knowledge of real world objects.
\item \textbf{Theoretical.} More general conceptual knowledge, implying cognitive constructions.
\item \textbf{Modeling.} The model is the formalized \emph{medium} of the scientific perspective, as diverse as Varenne's classifications of models functions~\cite{varenne2010simulations} (see below).
\item \textbf{Data.} Raw information that has been collected.
\item \textbf{Methods.} Generic structures of knowledge production.
\item \textbf{Tools.} Proto-methods (implementation of methods) and supports of others domains.
\end{itemize}

We choose to keep separate Methods and Tools, to insist on the support role of tools, and because development of both are related but not identical. The same way, Data domain and Empirical Domain are distinct, as new datasets do not systematically imply new knowledge of empirical facts. The Modeling Domain has a central role as we postulate that \emph{any knowledge on a complex system requires a model}.

\paragraph{Co-evolution of Knowledge}





We can now formulate the central hypothesis of our framework, that is partially contained in the positioning within Perspectivism. We postulate that \emph{any scientific knowledge construction on a complex system}\footnote{We believe that this intricate aspect of knowledge production is necessary present for Complex Systems, in echo of the remark on reflexivity in introduction. Even \emph{simple models} of complex systems do imply a conceptual complexity that requires complexity of knowledge to be grasped. This last assumption may be related to the nature of complexity and to the relation between computational complexity and complexity in the sense of weak emergence, that is suggested for example by \cite{2014arXiv1403.7686B} that explains emergence and decoherence from the quantum level by the NP-completude of fundamental equations resolution. These considerations are far beyond the reach of this paper, and we take as an assumption that complex systems necessitate complex knowledge, whereas simple knowledge (in the sense of non co-evolving domains and agents) \emph{can} exist for simple systems.} is a perspective in the sens of Giere. It is composed of knowledge contents within each domain, 
 that \emph{co-evolve} between themselves and with the other elements of the perspective, in particular the cognitive agents. The notion of co-evolution is taken in the sense of~\cite{holland2012signals}, i.e. of co-evolving entities being within strongly interdependent niches with circular causal relations and that have a certain independence with the exterior within their boundaries. 
We note the importance of weak emergence emergence in the sense of Bedau~\cite{bedau2002downward} in the construction of the perspective from the co-evolution of its components, as it corresponds to an autonomous upper level that can be understood alone, as the scientific knowledge can be. Note that a perspective does not necessarily have components in all domains, but should generally have in most.

\paragraph{Application}

\begin{figure}[h!]
\centering
\includegraphics[width=\textwidth]{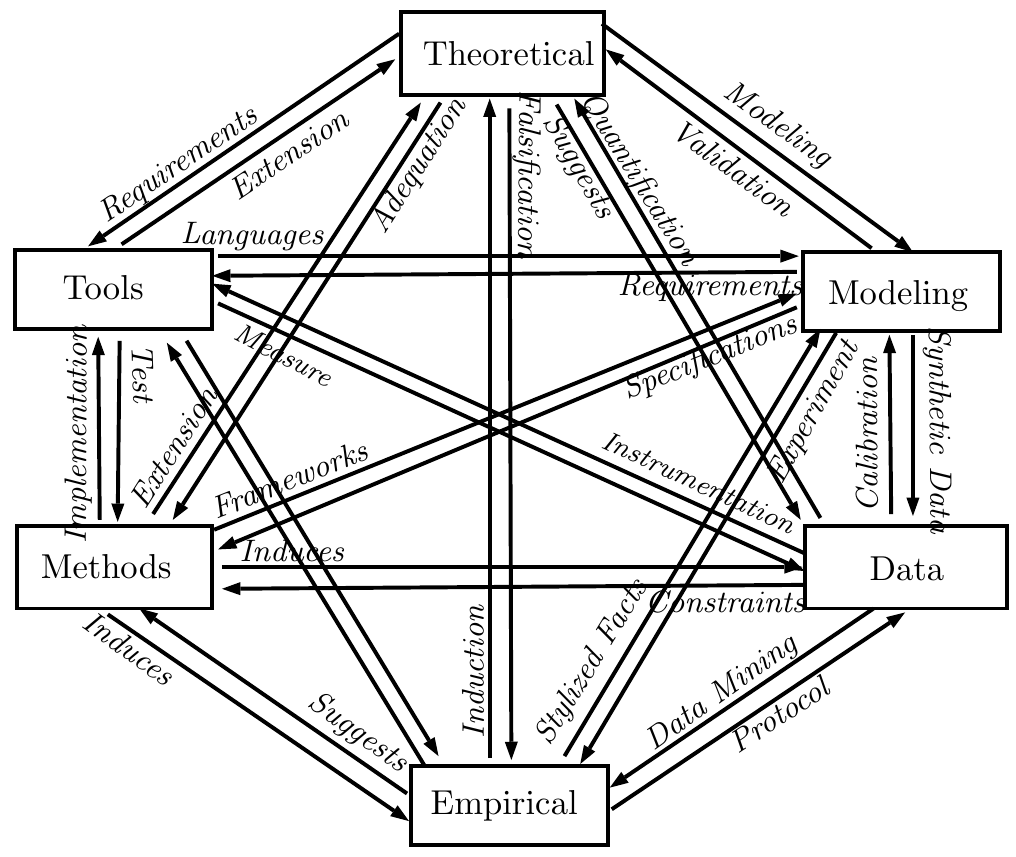}
\caption{\textbf{Projection of a perspective into a full network of knowledge domains.} To illustrate the domains and the interaction processes between them, we do the exercise of trying to qualify all possible binary relations between two given domains. This does not reflect the real structure of the framework, but is an aid to consider what interactions can be. Note that the nature of relations is not always the same here, some being constraints, other knowledge transfer, other processes within other domains such as synthetic data which is a methodology. This shows that some domains act as catalyzers for relations between others, in this network setting, what corresponds indeed to a situation of co-evolution.
}
\label{fig:fwk}
\end{figure}

The types of models to which our framework applies are supposed to be all possible models in a very loose sense, as Giere calls a model any medium of a perspective. A functional view of models as Varenne introduces~\cite{varenne2010simulations} (introducing a typology of models through functions, e.g. explicative models, simulation models, predictive models, comprehensive models, interactive models, etc.) is a way to grasp the variety. We can also see it in terms of more classical classifications, and apply it to mathematical, statistical, simulation, data or conceptuel models for example. Concerning the constraints given before, as all knowledge are co-evolving no domain is particularly privileged. No discipline either as these will have their different aspects be contained within the domains, and finally qualitative and quantitative methods are present and necessary in most. We show in Fig.~\ref{fig:fwk} a projection of knowledge domains as a complete network, to illustrate what relations between domains can be composed of.

\section{Discussion}

\subsection{Application Range}

We insist that our framework does not pretend to introduce a general epistemology of scientific knowledge, but far from that is rather targeted towards reflexivity in the understanding of complex systems. The level of generality is at a very different level, but the aim to practical implication in the handling of complexity contributes to a certain generic character in applications. It is furthermore particularly suited to study Complex Systems, since more reductionist approaches can handle more compartmented production of knowledge, whereas integration of disciplines and scales and therefore domains of knowledge has been emphasized as crucial to study complexity.

\subsection{Towards a formalisation}

The knowledge framework stays at an epistemological level, and its application could be formalized in a more systematic way. We give here a possible direction to achieve that, starting from the coupling of a formalization of the system model with one of the perspective. A perspective would be defined as a dataflow machine $M$ in the sense of~\cite{golden2012modeling} that gives a convenient way to represent it and to introduce timescales and data, to which is associated an ontology $O$ in the sense of~\cite{livet2010}, i.e. a set of elements each corresponds to an entity (which can be an object, an agent, a process, etc.) of the real world. Purpose and carrier of the perspective are contained in the ontology if they make sense for studying the system. Decomposing the ontology into atomic elements $O=(O_j)_j$ and introducing an order relation between ontology elements based on weak emergence ($O_j\succcurlyeq O_i$ if and only if $O_j$ weakly emerges of $0_i$) should yield a canonical decomposition of the perspective containing the structure of the system. The challenge would be then to link this decomposition with the canonical decomposition of the dataflow machine postulated by~\cite{golden2012modeling}, and then define knowledge domains within this coupling: data is in flows of the machine, modeling in the machine, empirical and theoretical in ontologies, methods in the structure of the tree. Such an enterprise with consistent operations is however totally beyond the scope of this paper, but would be a powerful development.

\vspace{-0.3cm}

\section{Conclusion}

\vspace{-0.3cm}

We have studied with mixed method the construction of a scientific theory in theoretical and quantitative geography, and from that inductively introduced a knowledge framework aiming at understanding the production of knowledge on complex system as a complex system itself, namely a perspective with co-evolving components within interdependent knowledge domains. Note that the approach is fully reflexive as several components were necessary. We believe our framework is a useful tool to study complexity and manage complex systems, since it explicits some choices and directions of developments that may otherwise be unconscious.

\vspace{-0.3cm}

\section*{Acknowledgments}

The author would like to thank D. Pumain and R. Reuillon for giving of their time for the interviews.

\vspace{-0.3cm}



\end{document}